\documentclass[usenatbib]{mn2e}

\usepackage{savesym}
\usepackage{amsmath}
\savesymbol{iint}
\usepackage{txfonts}
\usepackage{html}

\usepackage{subfigure}
\restoresymbol{TXF}{iint}

\usepackage{color}
\usepackage[normalem]{ulem}

\usepackage{subfigure}
\usepackage{graphicx}
\usepackage{epstopdf}
\usepackage{url}

\def\simless{\mathbin{\lower 3pt\hbox
{$\rlap{\raise 5pt\hbox{$\char'074$}}\mathchar"7218$}}}   %< or of order
\def\simmore{\mathbin{\lower 3pt\hbox
{$\rlap{\raise 5pt\hbox{$\char'076$}}\mathchar"7218$}}}   %> or of order
\newcommand{\be}{\begin{equation}}
\newcommand{\ee}{\end{equation}}
\newcommand       \bea          {\begin{eqnarray}}
\newcommand       \eea          {\end{eqnarray}}

\def\simlt{\mathrel{\hbox{\rlap{\hbox{\lower4pt\hbox{$\sim$}}}\hbox{$<$}}}}
\def\simgt{\mathrel{\hbox{\rlap{\hbox{\lower4pt\hbox{$\sim$}}}\hbox{$>$}}}}
\def\lesssim{\mathrel{\hbox{\rlap{\hbox{\lower4pt\hbox{$\sim$}}}\hbox{$<$}}}}

\def\gtrsim{\mathrel{\hbox{\rlap{\hbox{\lower4pt\hbox{$\sim$}}}\hbox{$>$}}}}

\title[Neutron precursors]{Neutron-powered precursors of kilonovae}

\author[]{Brian D.~Metzger$^{1}\thanks{E-mail: bmetzger@phys.columbia.edu}$, Andreas Bauswein$^{2}$, Stephane Goriely$^{3}$, Daniel Kasen$^{4,5}$\\
$^{1}$Columbia Astrophysics Laboratory, New York, NY, 10027, USA\\
$^{2}$Department of
Physics, Aristotle University of Thessaloniki, GR-54124 Thessaloniki,
Greece\\
$^{3}$Institut
d'Astronomie et d'Astrophysique, CP-226, Universit\'e Libre de Bruxelles,
1050 Brussels, Belgium\\
$^{4}$Nuclear Science Division, Lawrence Berkeley National Laboratory, Berkeley, CA, USA\\
$^{5}$Department of Physics and Department of Astronomy, University of California, Berkeley, CA, USA\\
}

\begin{document}
\date{Received / Accepted}
\pagerange{\pageref{firstpage}--\pageref{lastpage}} \pubyear{2014}

\maketitle

\label{firstpage}

\begin{abstract}

The merger of binary neutron stars (NSs) ejects a small quantity of neutron rich matter, the radioactive decay of which powers a day to week long thermal transient known as a kilonova.  Most of the ejecta remains sufficiently dense during its expansion that all neutrons are captured into nuclei during the $r$-process.  However, recent general relativistic merger simulations by Bauswein and collaborators show that a small fraction of the ejected mass (a few per cent, or $\sim 10^{-4}M_{\odot}$) expands sufficiently rapidly for most neutrons to avoid capture.  This matter originates from the shock-heated interface between the merging NSs.  Here we show that the $\beta$-decay of these free neutrons in the outermost ejecta powers a `precursor' to the main kilonova emission, which peaks on a timescale of $\sim$ few hours following merger at U-band magnitude $\sim 22$ (for an assumed distance of 200 Mpc).  The high luminosity and blue colors of the neutron precursor render it a potentially important counterpart to the gravitational wave source, that may encode valuable information on the properties of the merging binary (e.g.~NS-NS versus NS-black hole) and the NS equation of state.  Future work is necessary to assess the robustness of the fast moving ejecta and the survival of free neutrons in the face of neutrino absorptions, although the precursor properties are robust to a moderate amount of leptonization.  Our results provide additional motivation for short latency gravitational wave triggers and rapid follow-up searches with sensitive ground based telescopes.

\end{abstract} 
  

\section{Introduction} 
\label{sec:intro}

The final inspiral of stellar binaries containing neutron stars (NSs) and stellar mass black holes are the premier gravitational wave (GW) sources for the interferometric detectors Advanced LIGO and Virgo (\citealt{Abadie+10}).  A generic consequence of the merger process appears to be the ejection of neutron rich matter, either during the early, short-lived dynamical phase (e.g.~\citealt{Goriely+11b}) or in delayed outflows from the remnant accretion torus (e.g.~\citealt{Fernandez&Metzger13b}; \citealt{Perego+14}; \citealt{Just+14}).  NS binary mergers thus represent promising astrophysical sites for creating the heaviest elements in the universe via rapid neutron capture nucleosynthesis (the $r$-process; e.g.~\citealt{Lattimer&Schramm74}; \citealt{Freiburghaus+99}; \citealt{Goriely+11b}; \citealt{Rosswog+13}).  

The radioactive decay of unstable $r$-process nuclei provides a continuous energy source that keeps the ejecta hot even as it loses thermal energy to adiabatic expansion.  This powers a thermal optical/infrared transient with a peak luminosity approximately one thousand times brighter than a nova (a `kilonova') lasting for days to weeks after the merger (e.g.~\citealt{Li&Paczynski98}; \citealt{Metzger+10}).  Recent work suggests that matter containing even a small quantity of lanthanide or actinide elements possesses an opacity at optical frequencies that can exceed by orders of magnitude that of normal supernovae (\citealt{Kasen+13}; \citealt{Barnes&Kasen13}).  The sensitive dependence of the opacity on composition makes the light curves and colors of kilonovae sensitive diagnostics of physical processes during the merger and its aftermath (e.g.~\citealt{Metzger&Fernandez14}; \citealt{Grossman+14}).  

Kilonovae also represent a promising electromagnetic counterpart to the GW source (\citealt{Metzger&Berger12}).  Detecting such a counterpart would help identify the host galaxy of the merger, as will be essential to characterize the demographics of the merger population or to use them as cosmological probes (e.g.~\citealt{Holz&Hughes05}).  Among the many challenges to proposed GW-triggered follow-up programs are the faint luminosities of kilonovae and the poor sky localization provided by Advanced LIGO-Virgo (e.g.~\citealt{Kasliwal&Nissanke14}; \citealt{Singer+14}).  The high opacities of $r$-process nuclei also drive most of the emission to near infrared wavelengths, beyond the range where optical telescopes are most sensitive.  The identification of potential sources of more luminous bluer emission, even if shorter in duration, could greatly enhance the prospects for successful GW follow-up. 

In this paper we point out such an additional component of the kilonova emission that peaks at very early times, just hours after the merger, powered by the decay of free neutrons (\citealt{Kulkarni05}).  Most of the ejecta remains sufficiently dense as it expands that all free neutrons are captured into nuclei during the $r$-process.  Recent NS-NS merger simulations indicate, however, that a few per cent of the ejecta mass expands sufficiently rapidly that neutrons avoid capture.  Their subsequent $\beta$-decay in the outermost layers of the ejecta powers a short-lived `precursor' to the kilonova that may encode valuable information on the properties of the merging binary and on the NS equation of state (EOS).

This Letter is organized as follows.  In $\S\ref{sec:freeneutron}$ we describe the conditions for NS-NS mergers to produce a substantial abundance of free neutrons in the outermost ejecta.  In $\S\ref{sec:model}$ we describe a simplified one dimensional model for the free neutron precursor and present example light curves.  In $\S\ref{sec:discussion}$ we discuss our results and their implications for follow-up observations of GW-detected mergers.   

\vspace{-0.8cm}
\section{Free Neutrons in Merger Ejecta}
\label{sec:freeneutron}

Neutrons are so abundant in the dynamical ejecta from NS mergers that the nuclear flow is dominated by neutron captures and $\beta-$decays along the neutron drip line.  As heavy nuclei are synthesized, however, fission processes prohibit the formation of super-heavy nuclei and recycle the heavy matter into lighter fragments, which then restart capturing free neutrons in a process known as `fission cycling'.  Finally, when the neutron density decreases to $n_n \sim 10^{20}$ cm$^{-3}$, the timescale for neutron capture by the most abundant nuclei with closed-shell number $N = 126$ and 184 exceed a few seconds, after which time the nuclear flow is dominated by $\beta-$decays back to stability.    

Most of the ejecta expands to low densities over a characteristic timescale $\tau_{\rm exp} \sim 10-100$ ms, where $\tau_{\rm exp}$ is defined (somewhat arbitrarily) as the interval over which the density decreases below $4\times 10^{5}$ g cm$^{-3}$.  This relatively slow expansion results in a sufficiently high density that essentially all neutrons are captured into heavy nuclei, on a characteristic timescale of a second.  

\begin{figure}
\subfigure{
\includegraphics[width=0.47\textwidth]{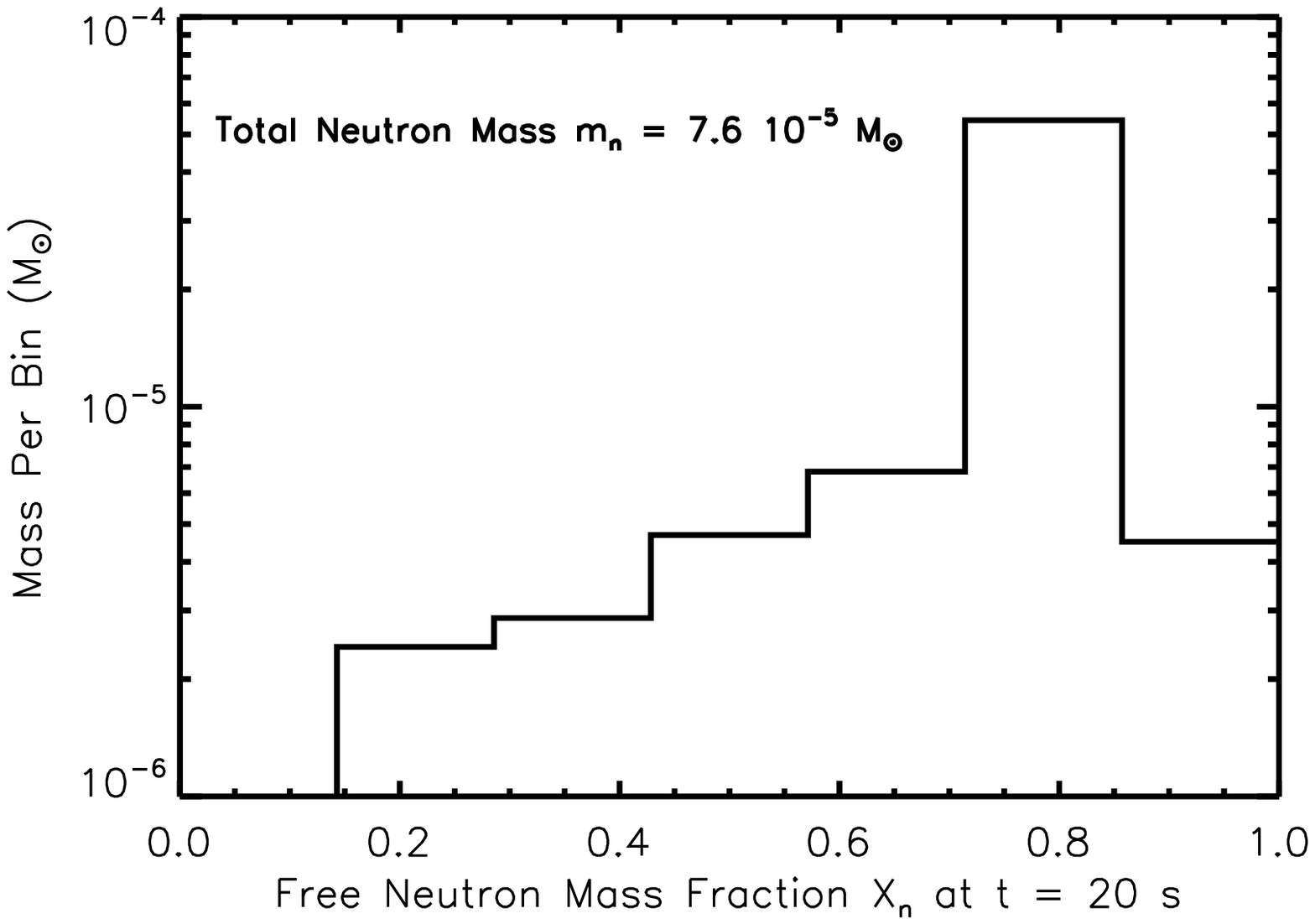}}
\subfigure{
\includegraphics[width=0.5\textwidth]{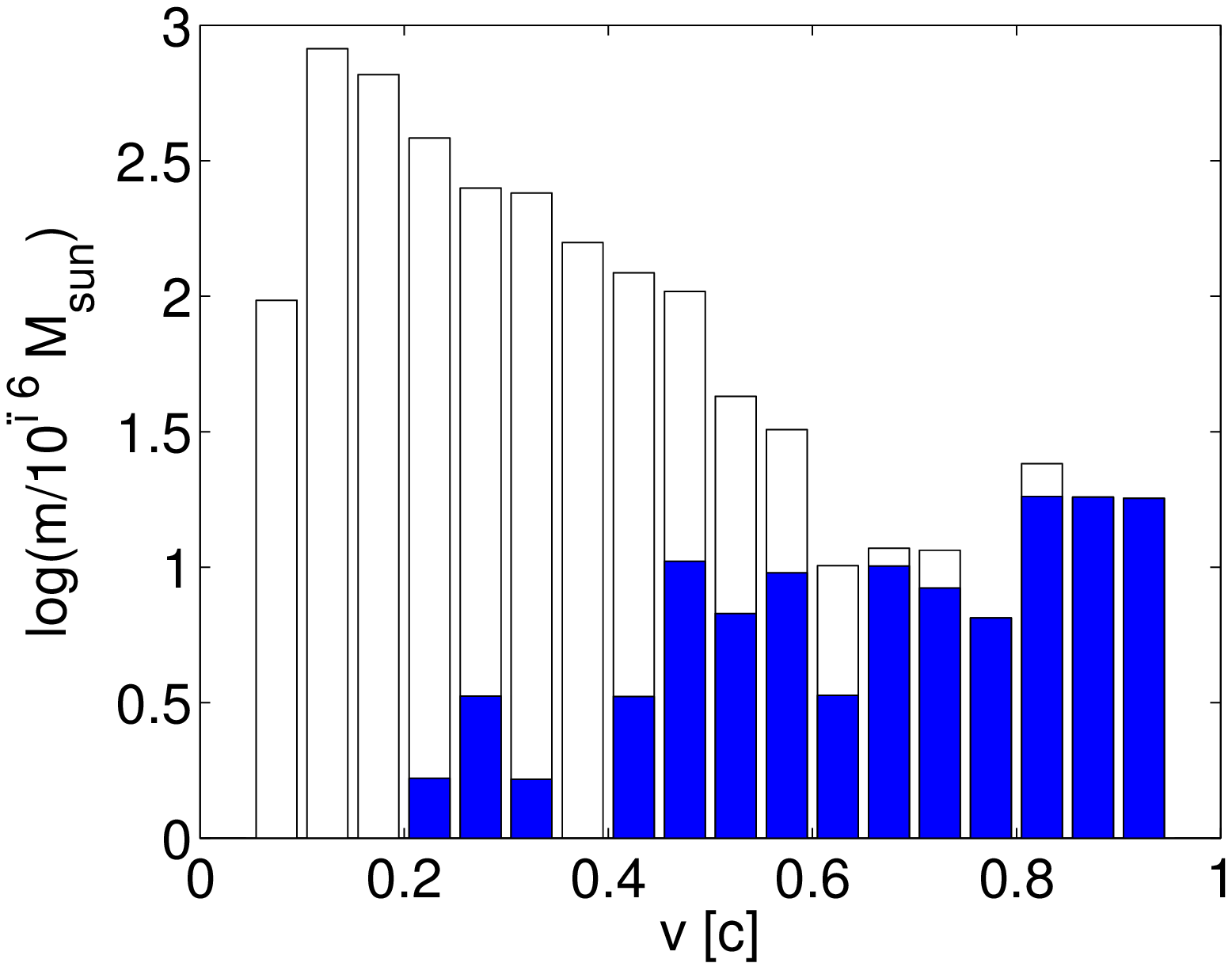}} 
\caption{{\it Top Panel:} Mass distribution of free neutron mass fraction $X_n$ of unbound fluid elements at t = 20 s post merger, calculated in \citet{Goriely+14} and \citet{Just+14} based on the 1.35-1.35 $M_{\odot}$ NS-NS merger simulation of \citet{Bauswein+13a} employing the DD2 EOS.  The total free neutron mass $\approx 7.6\times 10^{-5}M_{\odot}$ represents 2.5 per cent of the total ejecta mass $\sim 3\times 10^{-3}M_{\odot}$.  The histogram does not show the remaining 97.5 per cent of the ejecta for which the neutrons are entirely captured into nuclei ($X_{\rm n} \sim 0$).  {\it Bottom Panel:}  Velocity distribution of the ejecta for the same simulation at $t \approx 10$ ms.  White bars mark the total mass distribution, while blue bars mark just the matter that ends up rich in free neutrons ($X_{\rm n} > 0.1$).}
\label{fig:Bauswein}
\end{figure}

\citet{Goriely+14} and \citet{Just+14} recently discovered, however, that a small fraction of the ejecta in the merger simulations of \citet{Bauswein+13a} expands sufficiently rapidly ($\tau_{\rm exp} \lesssim$ 5 ms) for most neutrons to avoid being captured into nuclei, instead remaining asymptotically free.  This calculation was performed by using a nuclear reaction network to post-process SPH trajectories extracted from the 1.35-1.35$M_{\odot}$ NS binary merger simulation of \citet{Bauswein+13a} that employed the DD2 equation of state (\citealt{Typel+10}; \citealt{Hempel10}).  This fastest expanding matter originates from the shock-heated interface between the NSs, a component of the ejecta seen previously in other general relativistic (GR) simulations (e.g.~\citealt{Oechslin+07}; \citealt{Goriely+11b}; \citealt{Hotokezaka+13}).  Figure \ref{fig:Bauswein} (top panel) shows the mass distribution of the neutron mass fraction $X_{\rm n}$ at $t = 20$ s post merger, a time after the freeze-out of neutron captures but prior to the free neutron decay.  The bottom panel shows the velocity distribution of the ejecta ({\it white}), with those fluid elements rich in free neutrons ($X_{\rm n} > 0.1$) shown in blue. 

As mentioned above, given the relevant neutron capture cross sections, the requirement to retain a high mass fraction of free neutrons $X_n \sim 1$ is that the neutron density $n_{\rm n}$ decrease below $10^{20}$ cm$^{-3}$ ($\rho \sim 10^{-4}$ g cm$^{-3}$) on a timescale shorter than a few seconds.  If the ejecta were to expand homologously ($\rho \propto t^{-3}$) after a timescale $\gtrsim$ few $\tau_{\rm exp}$ from its initial density $\sim 10^{12}$ g cm$^{-3}$, then this condition translates into the density obeying $\rho \lesssim 10^{5}(\tau_{\rm exp}/{\rm ms})^{-3}$ g cm$^{-3}$ at the onset of homology.  This roughly motivates why the condition $\tau_{\rm exp} \lesssim$ few ms distinguishes those fluid elements with a high final neutron mass fraction.   Homogolous expansion is not achieved until a much later time $\sim 10-100$ s for the bulk of the ejecta with velocity $v \sim 0.1-0.2$ c, due to the fact that the ongoing energy input from radioactive decay ($\sim$ few MeV nucleon$^{-1}$) is comparable to the kinetic energy of the ejecta $\sim m_p v^{2}/2 \sim 4.7(v/0.1{\rm c})^{2}$ MeV nucleon$^{-1}$ (e.g., \citealt{Rosswog+14}).  However, radioactive energy input is relatively less important for the fastest ejecta ($v \gtrsim 0.5$ c), suggesting that homologous expansion may be achieved more rapidly.

\begin{figure}
\subfigure{
\includegraphics[width=0.4\textwidth]{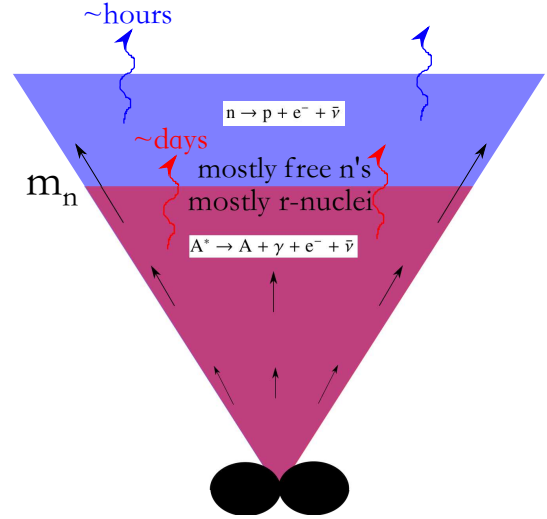}}
\caption{Schematic illustration of the ejecta from a binary NS merger leading to a neutron-powered precursor.  Matter ejected from the shock-heated interface between the merging NSs expands outwards with a range of velocities $v(m) \propto m^{-\beta}$ (eq.~[\ref{eq:mv}]; Fig.~\ref{fig:Bauswein}) where $\beta \approx 3$.  A small quantity of mass ejected first (mass coordinate $m < m_n \sim 10^{-4}M_{\odot}$; {\it blue}) expands sufficiently rapidly that neutrons avoid capture into nuclei, while for the bulk of the ejecta ($m > m_n$; {\it maroon}) neutrons are captured into nuclei during the $r$-process.  Energy released by the decay of free neutrons escapes the outer layers over a short timescale $t \lesssim $ hours, powering the precursor emission, which is relatively blue due to the high photosphere temperatures.  Radioactive energy released by $r$-process nuclei throughout the bulk of the ejecta (including that ejected by tidal tails and disk winds) escapes over a longer timescale $\sim$ days, with redder colors.} 
\label{fig:schematic}
\end{figure}

Though constituting only a few per cent of the dynamical ejecta by mass ($m_n \sim 10^{-4}M_{\odot}$) these fast expanding neutrons nevertheless have a dramatic impact on the kilonova emission because they reside in the outermost layers of the ejecta (Fig.~\ref{fig:schematic}).  The timescale for photon diffusion to this mass depth $\sim$ few hours (eq.~[\ref{eq:td0}]) is sufficiently short compared to the free neutron half-life of 10 minutes, that a substantial fraction of the energy released by neutron decay escapes as radiation.  

%Figure \ref{fig:schematic} is a schematic illustration of the ejecta geometry showing its relation to the precursor and main kilonova emission. 

Using 340,000 SPH particles to resolve the binary in this simulation, 106 SPH particles are found to be ejected with an expansion timescale sufficiently short that a substantial fraction of neutrons are not captured.  Very rapidly expanding fluid elements found in GR SPH simulations could in principle be numerical artifacts resulting from too few SPH particles; however, the smooth distributions of expansion timescale (Fig.~4 of \citealt{Just+14}) and the ejecta velocity (Fig.~\ref{fig:Bauswein}) argue against this possibility.  We note that \citet{Hotokezaka+13} have reported a head speed of the expanding ejecta of up to $\approx 0.8$ c, which is in good agreement with the distribution shown in Fig.~\ref{fig:Bauswein} (bottom panel).  Currently, it is unclear if similar trajectories occur in grid-based simulations and whether such calculations are able to resolve them if present.  Their absence in Newtonian SPH simulations might also be expected, because the strength of outflows from the merger interface region results in part from the deeper GR potential well.  Assuming the hydrodynamics is robust, the density profiles of fluid elements must be extrapolated to later times than the simulation duration of tens of milliseconds in order to follow the effects of neutron captures (\citealt{Goriely+11b}), a procedure upon which the final neutron fraction could be sensitive.  

 The other merger simulations presented in \citet{Bauswein+13a} show similar populations of rapidly expanding SPH particles, based on the same expansion timescale criterion $\tau_{\rm exp} \lesssim 5$ ms found to result in free neutrons in the case described above.  Asymmetric binaries appear to produce a greater quantity of fast expanding material; a $1.2-1.5M_{\odot}$ merger employing the DD2 EOS contains $\sim 2.6$ times as many particles satisfying $\tau_{\rm exp} < 5$ ms as the $1.35-1.35M_{\odot}$ case.  The quantity of fast-expanding fluid could also depend on the equation of state (EOS), because a softer EOS with a correspondingly smaller NS radius also produces stronger shock-heated outflows.  Simulations employing the very stiff NL3 EOS indeed produce only $\sim 1/3$ as many fast-expanding particles, while the softer SFHO EOS produces approximately the same number.  Post-processing of the trajectories with a reaction network finds a free neutron mass ranging from 0.6 to 5 per cent of the total ejecta mass for 13 different models.  Given the small numbers of SPH particles involved ($\sim 100$), these trends should be taken as only suggestive at this stage.

\vspace{-0.5cm}

\subsection{Shock Break-Out from Neutron Star Collisions}

Even if free neutrons are not abundant in the ejecta from standard binary NS mergers, other related physical scenarios could produce a small quantity of rapidly expanding ejecta.  Alternatively, such matter could be present even in standard NS mergers even if it is not well resolved by current simulations.  One example is shock break-out following the collision in a standard NS-NS merger (\citealt{Sekiguchi+11}; \citealt{Paschalidis+12}; \citealt{Kyutoku+14}) or as would be produced by the head-on collision of two NSs.  The latter could be physically realized if the binary eccentricity is stochastically raised due to Kozai oscillations by a tertiary companion (e.g.~\citealt{Katz&Dong12}).  

Two NSs that collide head-on at their escape velocity will dissipate an energy of $E_{\rm coll} \approx GM_{\rm ns}^{2}/2R_{\rm ns} \approx 3\times 10^{53}$ ergs, where $M_{\rm ns} \sim 1.4M_{\odot}$ and $R_{\rm ns} \approx 10$ km are the NS masses and radii, respectively.  The collision drives dual shocks outwards through each NS (e.g.~\citealt{Rosswog+09}) which as they approach the surface are accelerated to relativistic velocities by the decreasing density profile $\rho(r) \propto (R_{\rm ns}-r)^{n}$ (\citealt{Johnson&McKee71}), where $n=3$.  According to the formalism outlined by Nakar \& Sari (2011; their Sect.~2.1 and eqs.~[26]), each shell of mass $m = 10^{-4}m_{-4}M_{\odot}$ will be accelerated to a four velocity
\be
\beta\gamma = 0.93 m_{-4}^{-0.17}E_{53}^{0.62},
\ee
where $\gamma = (1-\beta^{-2})^{-1/2}$ and $E = E_{\rm coll}/2 = 10^{53}E_{53}$ erg is the total energy of each shock.  

Shock break-out thus naturally accelerates a small layer of mass $m \lesssim 10^{-4}M_{\odot}$ to mildly relativistic speeds, as pointed out by \citet{Kyutoku+14}.  If the thickness of the accelerated layer is comparable to the NS crust, $\Delta R \approx $ km, then its initial expansion time $\tau_{\rm exp} \approx \Delta R/7c \approx 5\times 10^{-7}$ s is indeed quite short.  Matter originating from the inner NS crust starts highly neutron rich (electron fraction $Y_e \lesssim 0.1$), but weak interactions such as neutrino absorptions and $e^{\pm}$ captures could in principle raise $Y_e$ substantially.  If the density of the shell following shock compression is 7 times higher than its initial value $\rho \approx m/4\pi R_{\rm ns}^{2}\Delta R$, then its post shock temperature is given by
\be
(4\pi/7)R_{\rm ns}^{2}\Delta R aT^{4} = m c^{2} \rightarrow kT \approx 52m_{-4}^{1/4}{\rm MeV}
\ee
The characteristic timescale for $e^{\pm}$ captures,
\be
\tau_{\pm} \simeq 2.1(T/{\rm MeV})^{-5}\,{\rm s} \approx 10^{-8}m_{-4}^{-5/4}\,{\rm s} 
\ee
is thus shorter than the expansion timescale for $m \gtrsim 10^{-5}M_{\odot}$.  If matter enters equilibrium with $e^{\pm}$ captures, then from Fig.~1 of \citet{Beloborodov03} we estimate that $Y_e$ could be raised to $\sim 0.3$ or above given the temperatures and densities corresponding to $m \approx 10^{-4}M_{\odot}$.  

A more detailed calculation of shock break-out, including both weak interactions and radiative losses, is necessarily to fully assess the properties of fast expanding, neutron-rich matter.  Such a calculation would also serve as a useful check on how the results of numerical simulations such as those presented in Figure \ref{fig:Bauswein} could be impacted by resolution of the neglect of weak interactions in the expanding matter.  
%For now we simply point out that $Y_e$ for the fastest expanding matter could be higher than calculated by current NS-NS merger simulations.  

\vspace{-0.7cm}

\section{Light Curve Model}
\label{sec:model}

As matter expands, its optical depth decreases.  The ejected mass has a gradient of velocities.  Following \citet{Piran+13}, we denote $m(v)$ as the mass with asymptotic velocity $> v$, which we parameterize as a single power law
\be
m(v) = \bar{m}\left(\frac{v}{\bar{v}}\right)^{-\beta}, \,\,v > \bar{v},
\label{eq:mv}
\ee
where $\bar{v} = 0.15$ c, $\beta = 3$, and $\bar{m} = 3\times 10^{-3}M_{\odot}$ (total ejecta mass) represent typical numbers for the dynamical ejecta from NS binary mergers.

The specific thermal energy $e(m)$ of each mass shell $[m,m+dm]$ evolves with time $t$ since the merger according to
\be
\frac{de}{dt} = -\frac{e}{t} + \dot{e} - \dot{e}_{\rm rad},
\label{eq:dedt}
\ee
where the first term accounts for energy loss via PdV work under radiation-dominated conditions.  The second term $\dot{e}  = \dot{e}_{\rm n} + \dot{e}_{\rm r}$ is the specific radioactive heating rate, where $\dot{e}_{\rm n} = 3.2\times 10^{14}X_{\rm n}(m,t)$ erg s$^{-1}$ g$^{-1}$ is the heating rate from the beta-decay of free neutrons \citep{Kulkarni05}, where $X_{\rm n}(m,t) = X_{\rm n,0}(m)\exp\left[-t/\tau_{\rm n}\right]$ (where $\tau_{\rm n} \approx 900$ s is the neutron decay timescale),  and $\dot{e}_{\rm r}$ is the heating rate from decaying $r$-process nuclei ($\beta$-decays and fission; e.g.~\citealt{Metzger+10}), which we model using the parameterization of \citet{Grossman+14}:
\be
\dot{e}_{\rm r} = 2\times 10^{18}(1-X_{\rm n,0})\left\{0.5 - \pi^{-1}\arctan\left[(t-t_0)/\sigma\right]\right\}\,{\rm erg\,s^{-1}\,g^{-1}},
\ee
where $t_0 = 1.3$ s and $\sigma = 0.11$ s.  The initial free neutron mass fraction $X_{\rm n,0}$ is assumed to vary with depth according to 
\be
X_{\rm n,0} = (1-2Y_e)(2/\pi)\arctan\left[m_{\rm n}/m\right],
\ee
where $Y_e$ is the electron fraction of the fastest ejecta.  This simple functional form accounts in a crude way for our expectation that at shallow depths $m \ll m_{\rm n}$ (small $\tau_{\rm exp}$) the ejecta contains mostly free neutrons, while at high depths $m \gg m_{\rm n}$ (large $\tau_{\rm exp}$) all free neutrons are captured into nuclei.  A transition mass $m_n \sim 10^{-4}M_{\odot}$ is expected based on the fraction $\sim$ few per cent of the dynamical ejecta that expands sufficiently rapidly to avoid neutron capture (Fig.~\ref{fig:Bauswein}).  Remaining mass not in protons or neutrons is assumed to be $r$-process nuclei, with a (temporally fixed) mass fraction $X_{r}(m) = 1-X_{\rm n,0}$.  The last term in equation (\ref{eq:dedt}), 
\be \dot{e}_{\rm rad}(m) = \frac{e}{(t_{\rm d} + t_{\rm cr})},
\label{eq:edotrad}
\ee
accounts for radiative losses, where
\be
t_{\rm d} = \frac{3m \kappa}{4\pi \beta vct}
\label{eq:td}
\ee
is the effective diffusion timescale, which is calculated in the optically thick limit from the diffusion equation for the energy flux $F = m\dot{e}_{\rm rad}/4\pi r^{2} = (c/3\kappa \rho)\partial [e\rho]/\partial r \approx c e \beta/3\kappa r $, where the radial gradient $\partial r =  v t/\beta$ is approximated from the velocity gradient (eq.~[\ref{eq:mv}]).  The term $t_{\rm cr} = R/c = t(c/v)$ limits the energy loss timescale to the light crossing time when $t_{\rm d} \ll t_{\rm cr}$.

The opacity $\kappa(m,t) = \kappa_{\rm es} + \kappa_{\rm r}$ is composed of the Thomson scattering opacity $\kappa_{\rm es} = (\sigma_T/m_p)(1-X_{\rm n}(m,t)-X_{r})$ of the electrons paired with the protons produced by neutron decay and the opacity of the $r$-process nuclei $\kappa_{\rm r}$.  The latter is dominated by line transitions of actinide and lanthanide elements and is approximately independent of the $r$-process mass fraction $X_{\rm r}$ (\citealt{Kasen+13}).  Although the $r$-process line opacity will in general be frequency-dependent and will change with temperature and ionization state of the ejecta, we approximate  it using a temperature-independent grey opacity $\kappa_{r} \approx 3-30$ cm$^{-2}$.  The highest value we consider, $\kappa_r = 30$ cm$^{2}$ g$^{-1}$, is close to the mean opacity predicted by extrapolating the results of \citet{Kasen+13} (their Fig.~10) to the higher interior temperatures few $\times 10^{4}$ K relevant to the peak of the precursor emission.  However, a lower opacity $\kappa_{r} = 3$ cm$^{2}$ g$^{-1}$ may be justified if weak interactions increase $Y_e$ sufficiently that the nuclear flow does not get past the N = 126 closed shell, as is necessary to reach the lanthanides.  Each d-shell species contributes an opacity around 0.1 cm$^{2}$ g$^{-1}$ and since lanthanide-free matter is a mixture of $\sim$ 30 d-shell elements, this results in a total opacity $\approx 3$ cm$^{2}$ g$^{-1}$.  
%Fortunately the gross properties of the precursor emission turn out to be relatively insensitive to the value of $\kappa_{r}$.    

Assuming blackbody radiation, the temperature of the thermal emission is given by $T_{\rm eff} \approx (L_{\rm tot}/4\pi \sigma R_{\rm ph}^{2})^{1/4}$, where $L_{\rm tot} = \int \dot{e}_{\rm rad}dm$ is the total luminosity (summing eq.~[\ref{eq:edotrad}] over all mass shells).  The radius of the photosphere $R_{\rm ph}$ at each time is approximated as the radius $R = v(m_{\rm ph})t$ of the mass shell $m_{\rm ph}$ for which the optical depth equals unity.  The flux at frequency $\nu$ and distance $D$ is given by
\be
F_{\nu}(t) \simeq \frac{2\pi h\nu^{3}}{c^{2}}\frac{1}{e^{h\nu/kT_{\rm eff}}-1}\frac{R_{\rm ph}^{2}}{D^{2}},
\ee
Note that each shell provides its maximum contribution to the light curve at the timescale defined by the condition $t = t_{\rm d}$ (eq.~[\ref{eq:td}]), 
\be
t_{\rm d,m} = \left(\frac{3m\kappa}{4\pi \beta v c}\right)^{1/2} \approx  3\,{\rm hr}\,\,\left(\frac{m}{10^{-4}M_{\odot}}\right)^{1/2}\left(\frac{\kappa}{10\,\rm cm^{2}\,g^{-1}}\right)^{1/2}\left(\frac{v}{0.5\,\rm c}\right)^{-1/2}.
\label{eq:td0}
\ee  

\vspace{-0.7cm}

\subsection{Results}

\begin{figure}
\subfigure{
\includegraphics[width=0.45\textwidth]{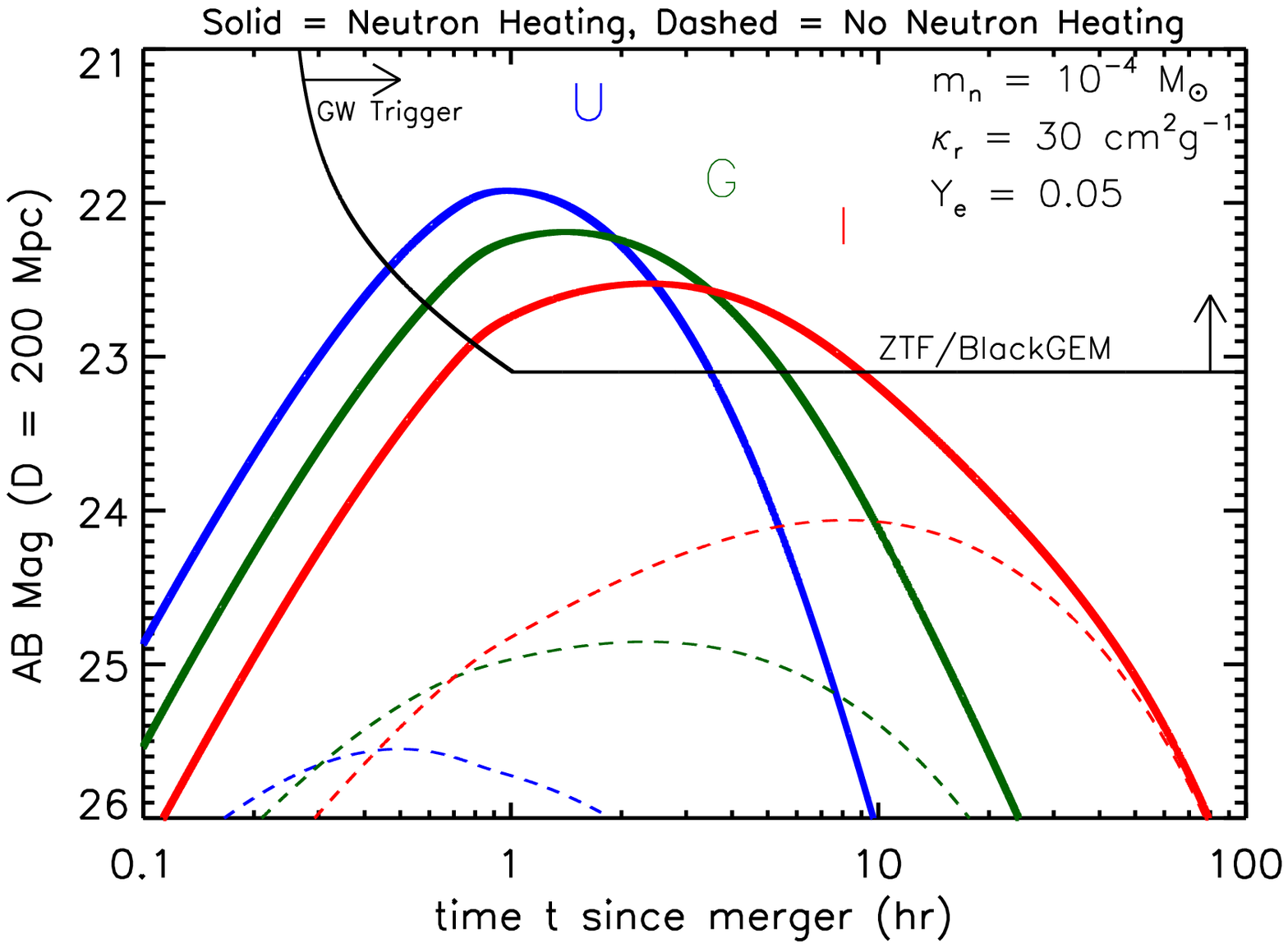}}
\subfigure{
\includegraphics[width=0.46\textwidth]{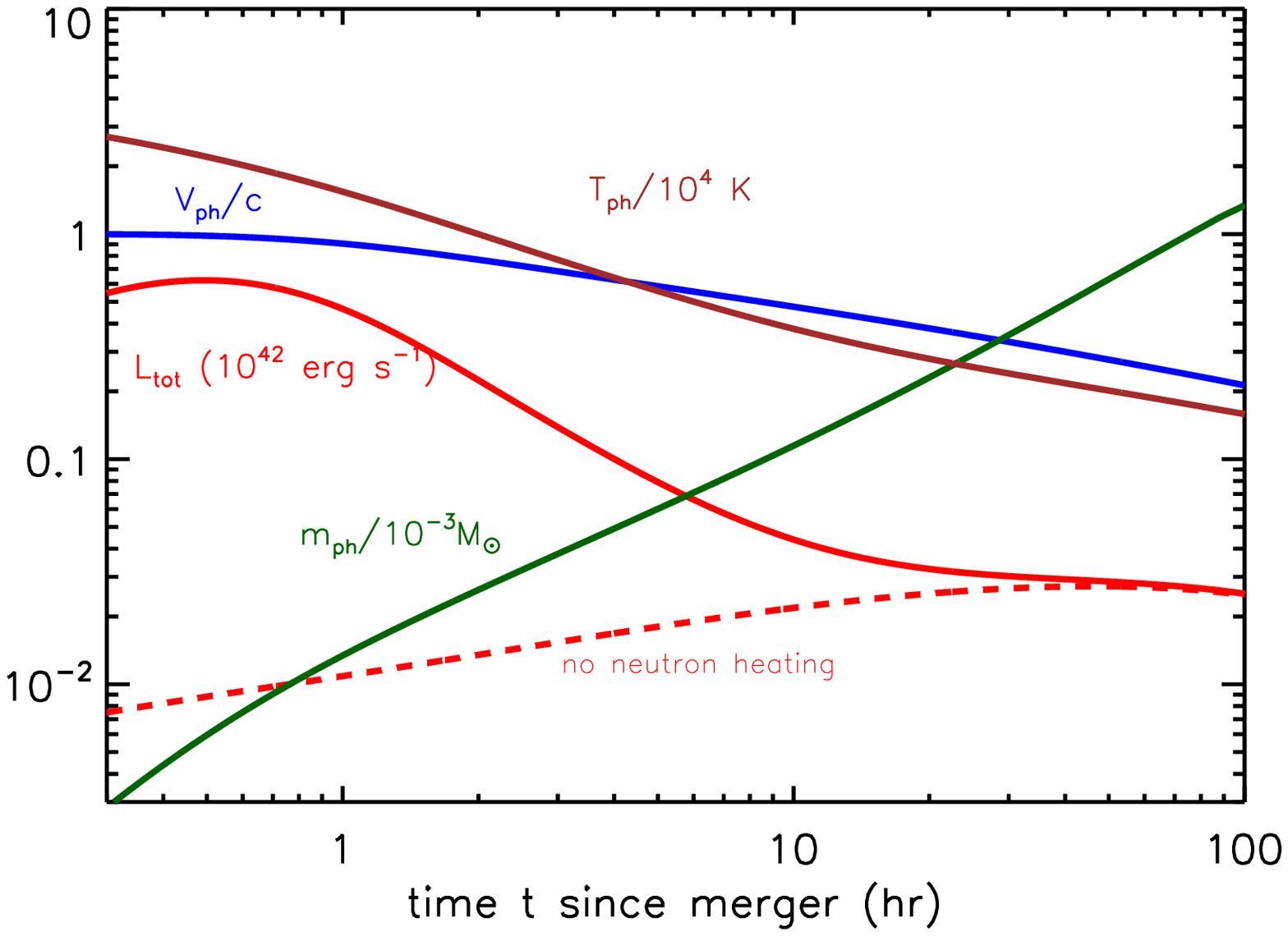}} 
\caption{{\it Top Panel:} Kilonova light curves, including the free neutron precursor, calculated for an assumed mass cut $m_{\rm n} = 10^{-4}M_{\odot}$ separating the ejected dominated by free neutrons and $r$-process nuclei.  AB magnitudes at U, G, and I bands are shown as a function of time after the merger for an assumed distance of 200 Mpc equal to the detection range of Advanced LIGO/Virgo for NS-NS mergers.  Solid lines show a model including the radioactive heating from free neutrons, while dashed lines show a case with neutron heating artificially set to zero.  Approximate sensitivity of the ZTF and BlackGEM telescope arrays are shown as a solid black line (see text).  {\it Bottom Panel:}  Bolometric luminosity $L_{\rm tot}$ and the mass $m_{\rm ph}$, radius $R_{\rm ph}$, and temperature $T_{\rm ph}$ of the photosphere, for the solution shown above.} 
\label{fig:LC1}
\end{figure}

Figure \ref{fig:LC1} (top panel) shows the multi-band light curve for our fiducial case corresponding to a neutron mass cut $m_{\rm n} = 10^{-4}M_{\odot}$, opacity $\kappa_r = 30$ cm$^{2}$ g$^{-1}$, electron fraction $Y_e = 0.05$, and an assumed distance $D = 200$ Mpc corresponding to the expected range of Advanced LIGO/Virgo for NS-NS mergers (\citealt{Abadie+10}).  Dashed lines show for comparison the light curves for the equivalent case with heating from neutron decay set to zero.  The contribution of neutron heating to the light curve is apparent as an excess at times $\lesssim$ few hours, which is most prevalent at U band due to the high photospheric temperature $T \sim 10^{4}$ K.  The bottom panel of Figure \ref{fig:LC1} shows the time evolution of the bolometric luminosity $L_{\rm tot}$ and the properties of the photosphere (radius, mass coordinate, and temperature).  The bolometric luminosity peaks at $\lesssim 10^{42}$ erg s$^{-1}$ on a timescale $\approx 0.5$ hr, radiated a total energy $\approx 2\times 10^{45}$ ergs that is only a few per cent of the total energy from the decaying neutrons, $m_{\rm n}\int \dot{e}_{\rm n}dt \approx 6\times 10^{46}(m_{\rm n}/10^{-4}M_{\odot})$ ergs.  

%Note that the I-band luminosity at late times near the peak of the main kilonova is overpredicted by this model since we assumed a fixed grey opacity for the $r$-process ejecta, which does not capture the substantial line blanketing at optical wavelengths (\citealt{Barnes&Kasen13}).  

Shown for comparison is the approximate sensitivity depth for the Zwicky Transient Facility (ZTF) and the BlackGEM array\footnote{\url{https://www.astro.ru.nl/wiki/research/blackgemarray}}.  We have assumed that ZTF can reach a depth g = 22.2 for a 600 s integration over its 47 deg$^{2}$ FoV (\citealt{Kasliwal&Nissanke14}), such that a LIGO/Virgo error region $\approx 100$ deg$^{2}$ could be covered with 2 pointings, which for a half hour integration each could achieve a sensitivity of $g \approx 23$ over a total time of $\sim$ 1 hour.  The BlackGEM array is more sensitive than ZTF (magnitude 23 for a 5 minute integration) but its smaller field of view results in a similar effective depth.  Flux sensitivity grows in time as $\propto t^{1/2}$, starting from the delay between the merger and the GW trigger, which we assume to be 15 minutes (\citealt{Cannon+12}).  

\begin{figure}
\subfigure{
\includegraphics[width=0.5\textwidth]{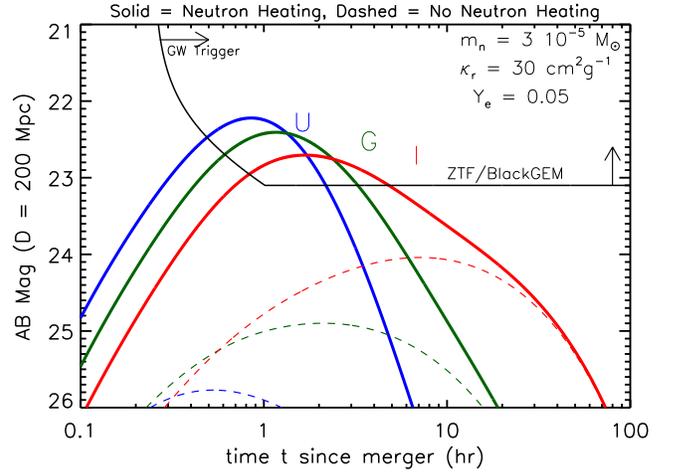}}
\caption{Same as top panel of Figure \ref{fig:LC1}, but calculated assuming a lower free neutron mass cut $m_{\rm n} = 3\times 10^{-5}M_{\odot}$.} 
\label{fig:LC2}
\end{figure}

\begin{figure}
\subfigure{
\includegraphics[width=0.5\textwidth]{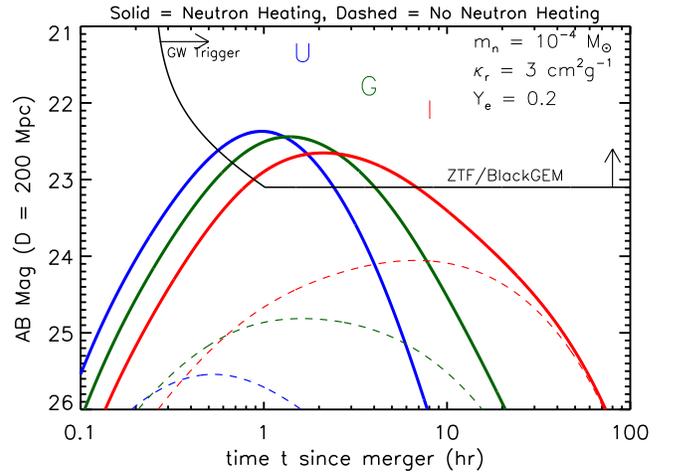}}
\caption{Same as top panel of Figure \ref{fig:LC1}, but calculated assuming a higher electron $Y_e = 0.2$ and (assuming lanthanides are not synthesized) a lower opacity $\kappa_{\rm r} = 3$ cm$^{2}$ g$^{-1}$.} 
\label{fig:LC3}
\end{figure}

Figure \ref{fig:LC2} shows a light curve similar to the fiducial case, but calculated assuming a lower mass cut $m_{\rm n} = 3\times 10^{-5}M_{\odot}$.  The lower mass cut reduces the U-band peak by $\approx 0.5$ mag.  The peak brightness does not depend on $m_{\rm n}$ as sensitively as would be naively expected based on the factor of 3 fewer neutrons.  This is because the outermost mass shells contribute a disproportionately large fraction of the early emission because they become transparent at an earlier time $t_{\rm d,m} \propto m^{1/2}$ (eq.~[\ref{eq:td0}]), resulting in a smaller fraction of the available energy from neutron decay $\propto m$ being lost to PdV work.  Figure \ref{fig:LC3} shows a case calculated for a lower opacity and higher electron fraction $Y_e = 0.2$, as would be appropriate if weak interactions increase the ejecta $Y_e$ sufficiently to avoid the production of lanthanide/actinide elements.  The peak brightness is comparable to the fiducial case because the lower free neutron mass $\propto 1-2Y_e$ is compensated by the lower opacity.
%\footnote{Note that the I-band luminosity at late times near the peak of the main kilonova is overpredicted by this model assuming that the bulk of the ejecta indeed produces lanthanide/actinide elements with high opacity}  

An early blue/visual bump thus appears to be a generic consequence of the fastest expanding ejecta being dominated by free neutrons.  We caveat that the specific color evolution will depend more sensitively on our assumption of a grey opacity.

%\begin{figure}
%\subfigure{
%\includegraphics[width=0.5\textwidth]{mag3.eps}}
%\caption{Same as top panel of Figure \ref{fig:LC1}, but calculated assuming a lower opacity $\kappa_r = 1$ cm$^{-2}$ g$^{-1}$ for the $r$-process opacity.  Note that this model overpredicts the late time emission on timescales of days because the opacity is expected to be much higher ($\kappa_r \gtrsim 10$ cm$^{2}$ g$^{-1}$) at these times (\citealt{Kasen+13}).
%} 
%\label{fig:LC3}
%\end{figure}
\vspace{-0.6cm}

\section{Discussion}
\label{sec:discussion}

\citet{Kulkarni05} first pointed out the potential importance that the energy released by decaying neutrons could have on the optical signatures of NS-NS mergers.  However, the vast majority of the ejecta remains sufficiently dense over the ensuing seconds that almost all neutrons are captured into heavy nuclei as a part of the $r$-process.  A possible exception emphasized here is the fast expanding matter ejected during the earliest phases of the merger from the interface between the surfaces of the merging NSs (\citealt{Just+14}).  This matter achieves low densities sufficiently rapidly to freeze-out a large abundance of neutrons $X_n \sim 1$ below a characteristic mass cut of $m_{\rm n} \sim 10^{-4}M_{\odot}$ (\citealt{Goriely+14}).  

Here we have shown even a small layer of free neutrons in the outermost ejecta can dramatically alter the early kilonova light curves.  This results primarily from the relatively long decay timescale of free neutrons as compared to the majority of the $r$-process nuclei, and the large quantity of energy released per decay.  The signature of free neutron decay is a blue/visual bump that peaks at a luminosity $L_{\rm tot} \sim$ few $10^{41}$ erg s$^{-1}$ on a timescale of $\lesssim$ few hours post merger, corresponding to a peak U-band magnitude $\sim 22$ at 200 Mpc.  Our work strongly motivates ongoing efforts to reduce the latency time of gravitational wave detectors (\citealt{Cannon+12}) and to develop rapid follow-up observing strategies with sensitive ground-based optical telescopes (e.g.~\citealt{Aasi+14}).

The neutron precursor is important for several reasons.  First, although the bulk of the radioactive energy comes out at later times from decaying $r$-process elements, the opacity of $r$-process matter is much higher than in normal supernovae \citep{Kasen+13}.  The resulting low photospheric temperature at late times produces emission that peaks in the far infrared (\citealt{Barnes&Kasen13}; although, see \citealt{Metzger&Fernandez14}) where follow-up telescopes are less sensitive.  By contrast, despite a similarly high opacity, the much higher photosphere temperatures of the precursor emission place the spectral peak squarely in the optical/ultraviolet range.  

Precursor emission could also provide a pristine probe of the fastest ejecta originating from the earliest stages of the merger.  Outflows from the merger interface which produce the fast expanding ejecta may be a common feature of all NS-NS mergers, but the quantity and electron fraction of this matter could depend sensitively on the NS EOS and the binary mass ratio.  For this reason the precursor could in principle be used to constrain these physical properties.  To the extent that the ejecta from black hole-NS mergers instead originates from the more slowly-expanding tidal tails, the presence or absence of the precursor could also help distinguish NS-BH from NS-NS mergers independent of the GW signal.  

Much additional work is necessary to confirm the predictions of the toy model presented here.  First, the robustness of the conclusion that free neutrons can escape the post merger environment should be explored using a wider range of numerical simulations and means for extrapolating the test particle trajectories to low densities.  The potential effects of weak interactions on the ejecta electron fraction must also be explored using merger calculations that include self-consistent neutrino transport, although preliminary work in this direction by \citet{Wanajo+14} finds that the outermost ejecta has $Y_e$ increased to only $\sim 0.1-0.2$.  The $r$-process line opacities should also be included in a more consistent way, accounting for their frequency and temperature dependence, the latter of which could vary significantly from the conditions experienced during the main kilonova phase at later times.  Finally, although we expect the ejecta to be relatively spherical on large scales, its true geometry is more complex than captured by a one dimensional model, motivating future multi-dimensional transport calculations.  

\vspace{-0.5cm}
\section*{Acknowledgments}
BDM acknowledges helpful conversations with Andrew Cumming, Thomas Janka, Oliver Just, Boaz Katz, and in particular Gabriel Mart\'inez-Pinedo.  BDM gratefully acknowledges support from the NSF grant AST-1410950 and the Alfred P. Sloan Foundation.  BDM and DK acknowledge support from the University of Washington Institute for Nuclear Theory workshop `The R-Process: Status and Challenges', where a portion of this work germinated.  AB is a Marie Curie Intra-European Fellow within the 7th European
Community Framework Programme (IEF 331873). Partial support for AB comes from``NewCompStar'', COST Action MP1304.  SG acknowledge financial support from FNRS (Belgium).  DK was supported by a Department of Energy Office of Nuclear
Physics Early Career Award, and by the Director, Office of Energy Research, Office of High Energy and Nuclear Physics, Divisions of Nuclear Physics, of the U.S. Department of Energy under Contract No. DE-AC02-05CH11231, and an NSF Division of Astronomical Sciences collaborative research grant AST-1206097.

\bibliographystyle{mn2e}
%\bibliography{../biblio/bibliography}
\bibliography{ms}

%\begin{thebibliography}{}
%\end{thebibliography}

\end{document}